\documentstyle [12pt]{article}

\def\be{\begin{equation}}
\def\ee{\end{equation}}
\def\bea{\begin{eqnarray}}
\def\eea{\end{eqnarray}}

\begin{document}

\begin{titlepage}
\title{Ortogonal rotation in the theory of finite-dimensional representations of quantum semisimple algebras.The case of $A_2$ algebra}
\author{A.N. Leznov\\
Universidad Autonoma del Estado de Morelos,\\ 
CCICAp,Cuernavaca, Mexico}

\maketitle

\begin{abstract}

The method of ortogonal rotations introduced in the previous papers of the author is used
for construction of the explicite form the generators of the simple roots for quantum (and ussual) semisimple algebras. All calculations are represented in explicit form for finite-dimensional representation $(p,q)$ of  $A^q_2$ algebra.

\end{abstract}

\end{titlepage}

\section{Introduction}

In the middle of the last century two famous papers on the theory of finite-dimensional representation of semisimple algebras appeared. Weyl \cite{Weyl} have found the explicit formula for characters of such representations for arbitrary semisimple series $A,B,C,D,E,F,G$. We will cite this result as a global aproach. In the paper of Gelfand-Zeitlin \cite{GZ} an explicit
formula was discovered for matrix elements of the generators but only for classical series $A,B,D$. During the second part of last century numerous unsuccesseful attempts were done to generalise infinitesimal approach of Gelfand-Zetlin to other Cartan series. 

In the papers of the author \cite{LEZ1},\cite{LEZ2} the solution of the problem was proposed on the level of solution of some system  of algebraic equations.  Examples for some representation were considered in details. The goal of the present paper is to generalise these results to arbitrary $(p,q)$ representation of semisimple quantum algebras of the rank 2 $A_2,B_2=C_2,G_2$.

In \cite{LEZ2} it was shown that these data are sufficient for construction of the generators of the finite-dimensional representation of all other semisimple algebras of arbitrary rank.  

\section{Preliminary comments and notations}

In this section we briefly present the results of \cite{LEZ1}. But the knowledge of the content of this paper is necessay for understanding the material below.
The equations defined the quantum algebra have the form 
\begin{equation}
[h_i X^{\pm}_j]=\pm K_{j,i} X^{\pm}_j,\quad [X^+_i,X^-_j]=\delta_{j,i}
{R_i- R_i^{-1}\over 2 \sinh w_i t},\quad R_i\equiv \exp w_it h_i\label{1}
\end{equation}

The first $2r$ equations (\ref{1}) really defines the selection rules for generators $X^{\pm}_i$
in the basis with the fixed proper values of the generators $h_i$.

These selection rules \cite{LEZ2} in the most simple way may be understood in terms of the diagramm of irreducible representation $(p,q)$ of one of the algebras the second rank $A_2,B_2=C_2,G_2$. Kartan matrices of these series are expressed in a similar form
$$
K=\pmatrix{ 2 & -w \cr
            -1 & 2 \cr}
$$
where $w=1,2,3$ correspondingly to $A_2,B_2=C_2,G_2$ (really Kartan matrix of $C_2$ is transport to Kartan matrix of $B_2$ above). And $w_i$ from the definition of the main equation of quantum algebras (\ref{1}) exactly coincides with $w_1=1,w_2=w$. 

We denote the point on the digramm of irreducible representation $(p,q)$ by two natural numbers indexes $k,s$ the number of the slating and number of vertical line correspondingly.
The object of investigatrion in \cite{LEZ2} is ortogonal matrices in each point of representation diagramm. The dimension of such matrices exactly coincides with multiplicity
of the basis vector with fixed proper values of the generators $h_1,h_2$ in corresponding point. The origin of the such construction in the structure of the irreducible representation of $A_1$ algebra (in further notation we do not do any difference between usual and quantum cases).
The irreducible representation of $A_1$ algebra is marked by index $P$ and each basis vector has natural index $j$. In canonical basis generator $h$ is diagonal and
$X^-$ is the lower triangular matrix with different from zero elements only under the main diagonal 
\be
(X^-)_{j+1,j}= \sqrt{{\{j+1\}\{P-j\}}} ,\quad \{j\}\equiv {\sinh jt},
$$
$$
X^+=(X^-)^T ,\quad (h)_{j,j}=P-2j,\quad 0\leq j \leq P \label{I}
\ee
Now let us consider the consequent action of the generator of some simple root. It connects in connection with the selection rules two point on diagramm with the multiplicity $N_i,N_{i+1}$
correspondingly. Thus it may be represented by $N_i\times N_{i+1}$ matrices which conserve its form under ortogonal rotations (in what follows we call such matrices as primitive matrix elements). If this operation will be continued sufficient number of times in "upper" direction ($X^+$) or in "dawn" one ($X^-$) we will come to zero result (representation is the finit one). Multiplicity on the boundary is always equal to unity. Thus the "upper" point on the boudary and the "dawn" one are connected with the help of some redducible representatin of $A_2$ algebra. Thus generators of some simple root is connected with the set of lines beggining on "upper" bondary and closed at the "dawn" one.  Each of these reducible representation are described by diagonal generators $h$ (with the correspoding multiplicity in each point) and upper (lower) triangular matrices with $N_i\times N_{i+1}$ blocks near the main diagonal. The whole dimension of such matrices is equal to $\sum N_i$. it is obvious that  this construction is invariant with respect to ortogonal rotations with the the ortogonal bloc matrices of the $N_i\times N_i$ dimension.    
In connection with the theory of representation of $A_1$ algebra each representation may be 
presented to canonical form with the help of ortogonal rotations. In other words the primitive matrix elements of the generator $X^-$ may be parametrised as follows
\be
(X^-_1)_{N_i,N_{i+1}}=O_{N_i} [Diag (N_i),0]O^{-1}_{N_{i+1}}\label{2}
\ee
where simbhol $[Diag (N_i),0]$ means $N_i\times N_{i+1}$ rechtangular matrix with different from zero elements on its "main" diagonal, which coinside with the matrix elements $(X^+)_{j,j+1}$
from (\ref{I}) ( don't mixed  meaning of this diagonality with the proper values of the generators $h_1$).

Now let us include into the game the second simple root $X^{\pm}_2$. The action of this generatores are denoted by slanting lines. Each point of the diagramm of representation is on the intersection of one vertical and one slanting line. About action of the generators 
$X^{\pm}_2$ and its primitive matrix elements it is possible repeat all said above with respect
to action of $X^{\pm}_1$ generator. 

The quantum numbers of the higest vector of representation 
denote by $(p,q),h_1=p,h_2=q$. By this reason vertical lines of the diagramm of representation will be called as $p$ lines, the slanting ones as a $q$ lines. We will assume that by ortogonal rotation the generators on the $p$ lines $X^{\pm}_1$ are passed to canonical  (diagonal in the sence above) form and generators $X^{\pm}_2$ parametrised similar to (\ref{2}). Thus
\be     
(X^-_1)_{(k,s|k+1,s)}=\lambda_p(k+1,s),\quad (X^-_2)_{(k,s|k,s+1)}=O_(k,s) 
\lambda_q(k,s+1)O^{-1}(k,s+1) \label{3}
\ee
The last determination is in the connection of our definition of coordinates on the plane of diagramm of representation given above. Under the motion on the $p$ lines changed by unity "coordinate" $k$, under the motion on $q$ lines changes coordinate $s$.

As it was explained above parametrization (\ref{3}) solves simultaneously two equations of quantum algebra (\ref{1}). Indeed in this case
\be
[X^+_1,X^-_1]={\sinh (th_1)\over \sinh t},\quad [X^+_2,X^-_2]={\sinh (twh_2)\over \sinh wt}
\label{4}
\ee
($w_1=1$!).
And for definition of ortogonal matrices it remains one unsolved up to now equation
$[X^+_2,X^-_1]=0$. Below we describe the trick with help of which this equation may be sucsessefully resolved.

With this aim let us consider four points of the representation diagramm with coordinates $(k,s),(k,s+1)(k+1,s)(k+1,s+1)$. The matrix elements of the commutator $[X^+_2,X^-_1]$ in connection with the selection rules may be different from zro only between the points $(k+1,s)$ and $(k,s+1)$. We have in a consequence
$$
(X^+_2)_{(k,s+1;k,s)}(X^-_1)_{(k,s;k+1,s)}-(X^-_1)_{(k,s+1;k+1,s+1)}(X^+_2)_{(k+1,s+1;k+1,s)}=0
$$
But $X^+_2=(X^-_2)^T$ and thus the last relation may be rewritten as $<k,s+1|X^+_2|k,s>=
<k,s+1|(X^-_2)^T|k,s>\equiv L(k,s)$:
\be  
L(k,s+1)\lambda_p(k+1,s)=\lambda_p(k+1,s+1)L(k+1,s+1)\label{5}
\ee
The last equation have the following structure. Unknown rechtangular matrix $L(k,s+1)$
multiplicates on the known diagonal matrice $\lambda_p(k+1,s)$ from the left equal to the product of known diagonal matrice $\lambda_p(k+1,s+1)$ and unown $L(k+1,s+1)$ in removed
($k\to k+1$) point. This equation in finite differences may be solved with similar (but not the same) results for all semisimple algebras of the second order. It gives the explicit dependence of unknown primitive matrix element $L(k,s)$ as function of coordinate $k$. 

Of course all consideration it is possible to repeat changing first and second simle roots by the places. Finally we come to the following system of equations for determining the ortogonal matrices in each point of the representation diagramm:  
\begin{equation}
(O^{s+1,k})\lambda_q^{s+1,k}(O^{s,k})^{-1}=L_{N(s+1,k),N(s,k)}\label{GRA1}
\end{equation}
\begin{equation}
(O^{s,k})^{-1}\lambda_p^{k+1,s}O^{s,k+1}=M_{N(s,k+1),N(s,k)}\label{GRA}
\end{equation}
where $N(s,k)$ are multiplicity in corresponding point of diagramma. The structure of 
rechtangular matrices $L,M$ will be defined below.

We would like to emphasize that the equations (\ref{GRA}), (\ref{GRA1}) are equivalent to the following ones $[X^+_1,X^-_1]={\sinh th_1\over \sinh t}, [X^+_2,X^-_2]={\sinh twh_2\over \sinh wt}, [X^+_1,X^-_2]=0$ defining the quantum semisimple algebra of the second rank.  

\section{The $(p,q)$ reprsentation of $A_2$ algebra}

The representation diagramm in this case has a hegsagon form. It has six vertexes.
The selection rules for simple roots generator $X^-_{1,2}$ are consequently the following ones 
$$
\Delta h_1=-2,\quad \Delta h_2=1;\quad \Delta h_1=1,\quad \Delta h_2=-2 
$$
(from the point of the higest vector to all other points of diagramm it is possible to go with help of consequent action of the generators with negative indexes).

Using these selection rules it is possible to reconsruct the upper boundary of the representation diagramm. It consists from the points
$$
(p,q),(p+1,q-2)..,(p+s,q-2s)..(p+q,-q); 
$$
$$
(p+q-1,-(q+1)),..(p+q-k,-(q+k)),  (q,-(p+q))
$$
and all other points of the diagramm can be obtain by corresponding number of mooving along
the vertical $p$ lines.

The maximal values of indexes of representations of $A_1$ algebra of the first simple root counting from the point of the higest vector $(p,q)$ are the following ones $p,p+1,..,p+s,..p+q; p+q-1,..q)$. The maximal values of representations indexes of slating $q$ lines connected with the second simple root are the following one $(q,q+1,....q+p;q+p-1,...,p)$. 

The problem of multiplicite in the case of $A_2$ is solved in the following way.
On $(p,q)$ diagramm it is necessary to indicate the points corresponding to extra bondary of the irreducible representations $(p,q-1),(p,q-2),....(p,0)$ ( we assume that $q\leq p$,this is not essential because representations $(p,q)$ and $(q,p)$ of $A_2$ algebra are equivalent). 
The multiplicity of the points on the boundary of the diagramm of the $(p,q-l)$ representation
are the same and equal to $l+1$. Multiplicity on all points of the diagramm $(p,0)$ representation (this diagramm has the form of triangular) is the same and equal $q$. 

By the vertical line connected points $(p+q,-q)$ and $(-(q+p),p)$ hegsogen diagramm is devided on two trapeciums- left and right in what follows.     

\subsection{Situation inside of the left trapecium}

Now we would like to come back to the general formulae (\ref{GRA}). In the notation above the main equations of \cite{LEZ2} may be rewritten in the following form (in the left
part of diagramm). This equation equivalent to conditions of commutativity $[X^+_1,X^-_2]=0$
simple positive and negative roots with different indexes.
\be
\lambda_p^{k+1,s+1}L^{k+1}_{s+2,s+1}=L^k_{s+2,s+1}\lambda_p^{k+1,s} \label{MNE}
\ee
In rewriting (\ref{5}) in the last form we take into account that multiplicity in the point
$(k,s), k\leq s$ is equal $s+1$ as it was described above. The index of $A_1$ representation 
of $s$ line equal $p+s$ and $p+s+1$ on $s+1$ vertical line. Thus for matrix elements of the diagonal matrix $\lambda_p^{k+1,s+1}$ from (\ref{1}) we have  
$\lambda_i=\sqrt{{\{k+2-i\}\{p+s-k+2-i\}\over \{1\}^2}}$ and $\lambda_p^{k+1,s}$ is the same with $\lambda_i=\sqrt{{\{k+2-i\}\{p+s-k+1-i\}\over \{1\}^2}}$.
Let us consider the $(1,1)$ term of the last matrix equation:
$$
\sqrt{{\{k+1\}\{p+s-k+1\}\over \{1\}^2}}(L^{k+1}_{s+2,s+1})_{1,1}=
(L^k_{s+2,s+1})_{1,1} \sqrt{ {\{k+1\}\{p+s-k\}\over \{1\}^2}}
$$
from which we conclude 
$$
(L^k_{s+2,s+1})_{1,1}=a_1(s)\sqrt{\{p+s-k+1\}}
$$
Let us consider $(2,1)$ matrix element of the same equation
$$
\sqrt{{\{k\}\{p+s-k\}\over \{1\}^2}}(L^{k+1}_{s+2,s+1})_{2,1}=
(L^k_{s+2,s+1})_{2,1} \sqrt{ {\{k+1\}\{p+s-k\}\over \{1\}^2}}
$$
from which we conclude 
$$ 
(L^k_{s+2,s+1})_{2,1}=b_1(s)\sqrt{\{k\}}
$$
Continuing such consideration we conclude that in the case $k\neq s$ the rechtangular $(s+2)\times (s+1)$ matrix $L^k$ has the different from zero elements only on its main diagonal and under it 
$$
(L^k_{s+2,s+1})_{i,i}=a_i\sqrt{\{p+s-k+2-i\}} \quad 1\leq i\leq s+1,
$$
$$
(L^k_{s+2,s+1})_{i+1,i}=b_i\sqrt{\{k+1-i\}}
$$
where all $a_i,b_i$ are the functions of only one parameter $s$.

Now (\ref{GRA}) becomes the equation for definition of the ortogonal matrices $O^s$ and
parameters $(a_i(s),b_i(s)$ which define prinitive matrix elements of $A_2$ algebra.
And we going to solve this problema.

First of all let us use the fact that from (\ref{GRA}) it is known that $L_{s+1,s+2}$ is the linear combination of the $s+1$ first columns of ortohonal $(s+2)\times (s+2)$ matrix $O^{s+1,k}$. This means that its $(s+2)$ column may defined from calculations of the minores of $(s+1)$ order of the matrix (\ref{YX}) and deviding the result on the product of the roots of diagonal matrix $\lambda_q^{s+1,k}$ from (\ref{GRA1}). We emphasing that the knowledges of the explicit expressions for coefficient $a_i,b_i$ it is not necessary for this calculations and more other the result will be used for their definition. Fullfiling this opration we obtain functional dependence of the last $(s+2)$ column on parameters $(k,s)$
$$
O^{s+2}_{1,s+2}=l_1\sqrt{ {\{k\}\{k-1\}....\{k-s\} \over \{q+k-s\}...\{q+k-2s\}}},
$$
$$
O^{s+2}_{2,s+2}=l_2\sqrt{ {\{p-k+s+1\}\{k-1\}....\{k-s\} \over \{q+k-s\}...\{q+k-2s\}}}
$$
$$
O^{s+2}_{3,s+2}=l_3\sqrt{ {\{p+s-k+1\}\{p+s-k\}\{k-2\}....\{k-s\} \over \{q+k-s\}...\{q+k-2s\}}}
$$
$$
.........................................................................................
$$
$$
O^{s+2}_{s+2,s+2}=l_{s+2}\sqrt{ {\{p+s-k+1\}....\{p-k+1\} \over \{q+k-s\}...\{q+k-2s\}}}
$$
The condition of ortogonality leads to a system of a linear equations for definition numeracal parameters $l_i$. The result present below:  
$$
l_1=\sqrt{{\{p+q+1\}...\{p+q-s+1\} \over \{p+s+1\}...\{p+1\}}}
$$
$$
l_2=\sqrt { {\{s+1\}\{q-s\}\{p+q\}...\{p+q-s+1\} \over \{1\}\{p+s+1\}\{p+s-1\}..\{p\}}}
$$
$$
l_3=\sqrt { {\{s+1\}\{s\}\{q-s\}\{q-s+1\}\{p+q-1\}...\{p+q-s+1\} \over \{1\}\{2\}\{p+s\}
\{p+s-1\}\{p+s-3\}..\{p-1\}}}
$$
$$
.........................................................................................
$$
$$
l_{s+2}=\sqrt{ { \{q\}\{q-1\}......\{q-s\} \over \{p+1\}...\{p-s+1\}}}
$$
Useffull relations follows from the definition $l_k$
$$
{l_i\over l_{i+1}}=\sqrt{{\{i\}\{p+q+2-i\}\{p+1-i\}\{p+s+4-2i\}\{\over \{p+s+2-2i\}\{p+s+3
-i\}\{s+2-i\}\{q-s+i-1\}}}
$$
As it was mentioned above $(s+2)\times (s+1)$ matrix $L$ is the linear combination of the
$(s+1)$ columns of ortohonal $(s+2)\times (s+2)$ matrix $O^{s+1,k}$. By this reason each column of the matrix $L$ must be ortogonal to $(s+2)$-th column. The components of which were obtained above. This fact leads to one local equation connected $a,b$ functions:
\be
a_i l_i+b_i l_{i+1}=0 \label{LK1}
\ee

Now we would like to show that by similar consideration it is possible to find the elements of the first column  $O^{s+2}_{i,1}$. For this aim let us rewrite the main equation (\ref{GRA1}) in equivalent form
\be
O^{s+1,k})\lambda_q^{s+1,k}=L^k_{s+2,s+1}O^{s,k}\label{!}
\ee

From the last equation it is easy by induction obtain the following dependence matrices elements of the first column from the parameters $(s,k)$ $(\lambda^q_1=\sqrt{{\{s+1\}\{q+k-s\}\over \{1\}^2}})$: 
$$
O^{s+2}_{1,1}=m_1\sqrt{ {\{p-k+s+1\}\{p-k+s\}....\{p-k+1\} \over \{q+k\}...\{q+k-s\}}},
$$
$$
O^{s+2}_{2,1}=m_2\sqrt{ {\{k\}\{p-k+s\}....\{p-k+1\} \over \{q+k\}...\{q+k-s\}}}
$$
$$
O^{s+2}_{3,s+2}=m_3\sqrt{ {\{k\}\{k-1\}\{p-k-1\}....\{p-k+1\} \over \{q+k\}...\{q+k-s\}}}
$$
$$
.........................................................................................
$$
$$
O^{s+2}_{s+2,1}=m_{s+2}\sqrt{ {\{k\}....\{k-s\} \over \{q+k\}...\{q+k-s\}}}
$$
There are no difficulties in determination of numerical parameters $m_i$. 
$$
m_1=\sqrt{{\{q\}...\{q-s\} \over \{p+s+1\}...\{p+1\}}}
$$
$$
m_2=-\sqrt { {\{s+1\}\{p+q+1\}\{q\}...\{q-s+1\} \over \{1\}\{p+s+1\}\{p+s-1\}..\{p\}}}
$$
$$
m_3=\sqrt { {\{s+1\}\{s\}\{p+q+1\}\{p+q\}\{q\}....\{q-s+2\} \over \{1\}\{2\}\{p+s\}
\{p+s-1\}\{p+s-3\}..\{p-1\}}}
$$
$$
.........................................................................................
$$
$$
m_{s+2}=(-1)^{s+1}\sqrt{ { \{p+q+1\}\{p+q\}......\{p+q-s+1\} \over \{p+1\}...\{p-s+1\}}}
$$
The following relations are direct consequence of above the definition $m_k$
$$
{m_i\over m_{i+1}}=-\sqrt{{\{i\}\{q-1+i-1\}\{p+1-i\}\{p+s+4-2i\}\{\over \{p+s+2-2i\} 
\{p+s+3-i\}\{s+2-i\}\{p+q+2-i\}}}
$$

Now calculating first column of the equation (\ref{!}) we obtain 
\be
\sqrt{{\{s+1\} \over \{1\}^2}}m_i=b_{i-1}m^-_{i-1}+a_im^-_i\label{LK2} 
\ee
Together with (\ref{LK1}) we have two equation which allow determine dependence of the parameters $a_i,b_i$ from $s$ coordinate. The result is the following 
$$
a_i=\sqrt{ {\{s+2-i\}\{q-s+i-1\}\{p+s+3-i\}\over \{1\}^2\{p+s+3-2i\} \{p+s+2-2i\}}},
$$
$$ 
b_i=-\sqrt{ {\{i\}\{p+q+2-i\}\{p+1-i\}\over \{1\}^2\{p+s+3-2i\} \{p+s+4-2i\}}}
$$
Instead of equation of equation (\ref{LK1}) it is possible to use equation similar to (\ref{LK2})  
\be
\sqrt{{\{s+1\} \over \{1\}^2}}l_i=b_{i-1}l^-_{i-1}+a_il^-_i\label{LK3} 
\ee
and have no deals with equation in finite differences but result of course conserves its form given above.

\subsection{Situation inside of the right trapezium}

As it was mentioned above (we are working with representation $(q\leq p)$) the multiplicities have regular character and change on unity when $k\to (k+1)$ under the going dawn on the $p$ line of diagramm of representation up to the value $k=q$. After this they are not changed and
equal to $q+1$ (up to symmetrical to $k=q$ point of diagramm after which they decreases on unity
with each next step).
  
On the upper boundary the proper values of $h_1,h_2$ are the following $h_1=p+2q-s,h_2=-s$.
As in the case of the left trapezium consider four points with coordinates $(s,k),(s+1,k),(s+1,k+1),(s,k+1)$ and equation similar to (\ref{MNE}). It is only necessary to keep in mind that reducible representation of $A_1$ algebra begin from index $p+2q-s$ (instead of $(p+s)$ in (\ref{MNE})) and now the number of steps from the upper boundary $\delta$ up to intersection of s-th $p$ line with k-th $q$ line is determined from the equation $q+k-2s=-s+\delta$ equal to $q+k-s=\delta$.
On this way we obtain the following expression for non zero elements of $L$ matrix: 
\be
(L^k_{s+2,s+1})_{i,i}=a_i\sqrt{\{p+q-k+2-i\}} \quad 1\leq i\leq s+1,
$$
$$
(L^k_{s+2,s+1})_{i+1,i}=b_i\sqrt{\{q+k+1-s-i\}}\label{TRII}
\ee
with additional condition $a_1=0$. Thus the left upper corner of the matrix $L$
begins from the term $b_1\sqrt{\{q+k+1-s-i\}}$ ($1\leq i \leq (q+1)$).
Using equation (\ref{!}) with matrix $L$ above we come to the following explicit expression
for components of $(q+1)\times (q+1)$ ortogonal matrix in point (k,s) ($q+1\leq s$)
$$
O^{s,k}_{1,1}=m_1\sqrt{ {\{p+q-k\}....\{p-k+1\} \over \{q+k\}...\{k+1\}}},
$$
$$
O^{s,k}_{2,1}=m_2\sqrt{ {\{k+q-s\}\{p+q-k-1\}....\{p-k+1\} \over \{q+k\}...\{k+1\} }}
$$
$$
O^{s,k}_{3,1}=m_3\sqrt{ {\{k-s+q\}\{k-s+q-1\}\{p-k+q-2\}....\{p-k+1\} \over \{q+k\}...\{k+1\}}}
$$
$$
.........................................................................................
$$
$$
O^{s,k}_{q+1,1}=m_{q+1}\sqrt{ {\{k-s+q\}....\{k-s+1\} \over \{q+k\}...\{k+1\}}}
$$
There are no difficulties in determination of numerical parameters $m_i$. 
$$
m_1=\sqrt{{\{s\}\{s-1\}....\{s-q+1\} \over \{p+2q-s\}...\{p+q-s+1\}}}
$$
$$
m_2=-\sqrt { {\{q\}\{p+q+1\}\{s\}....\{s-q+2\} \over \{1\}\{p+2q-s\}\{p+2q-s-2\}...\{p+q-s\}}}
$$
$$
m_3=\sqrt { {\{s+1\}\{s\}\{p+q+1\}\{p+q\}\{q\}....\{q-s+2\} \over \{1\}\{2\}\{p+s\}
\{p+s-1\}\{p+s-3\}..\{p-1\}}}
$$
$$
.........................................................................................
$$
$$
m_{q+1}=(-1)^{q+1}\sqrt{ { \{p+q+1\}\{p+q\}......\{p+2\} \over \{p+q-s+1\}...\{p-s+2\}}}
$$
Absolutely by the same arguments it is possible to find the components of the last column 
$$
O^{s+1}_{1,q+1}=l_1\sqrt{ {\{k-q+1\}\{k-q\}....\{k-s\} \over \{k+1\}...\{k-s+q\}}},
$$
$$
O^{s+1}_{2,q+1}=l_2\sqrt{ {\{p+q-k\}\{k-q+1\}....\{k-s\} \over \{k+1\}...\{k-s+q\}\{k-s+q-1\}}}
$$
$$
O^{s+1}_{3,q+1}=l_3\sqrt{ {\{p+q-k+1\}\{p+q-k-1\}\{k-q+1\}....\{k-s\} \over \{k+1\}...\{k-s+q-1\}\{k-s+q-2\}}}
$$
$$
.........................................................................................
$$
$$
O^{s+1}_{q,q+1}=l_q\sqrt{ {\{p+q-k\}....\{p-k+2\} \{k-s\} \over \{k+1\}...\{k-q+2\}}}
$$
$$
O^{s+1}_{q+1,q+1}=l_{q+1}\sqrt{ {\{p+q-k\}....\{p-k+2\} \{p-k+1\} \over \{k+1\}...\{k-q+2\}}}
$$
The condition of ortogonality leads to a system of a linear equations for definition numeracal parameters $l_i$. The result present below:  
$$
l_1=\sqrt{{\{p+q+1\}...\{p+2q-s\} \over \{p+1\}...\{p+q-s\}}}
$$
$$
l_2=\sqrt { {\{s+2-q\}\{q\}\{p+q\}...\{p+2q-s\}\{p+2q-s-2\} \over \{1\}\{p+1\}....
\{p+q-s-1\}}}
$$
$$
l_3=\sqrt { {\{q\}\{q-1\}\{s+2-q\}\{s+3-q\}\{p+q-1\}...\{p+2q-s-1\}\{p+2q-s-4\} \over \{1\}\{2\}\{p+1\}\{p\}..\{p+q-s-2\}}}
$$
$$
.........................................................................................
$$
$$
l_{q+1}=\sqrt{{\{s+1\}\{s\}......\{s-q+2\} \over \{p+q-s\}...\{p-s+1\}}}
$$

Absolutely by the same way as in previous subsection we obtain two equations
$$
\sqrt{{\{s+1-q\}\over \{1\}^2}}\{k-s\}l_i=\tilde l_i\{q+k-s+1-i\}b_i+\tilde l_{i+1}\{p+q-k+2-i\}
a_{i+1} 
$$
$$
\sqrt{{\{s+1\}\over \{1\}^2}}l_i=\tilde m_i b_i+\tilde m_{i+1}a_{i+1} 
$$
where as always $\tilde m_i(s)\equiv m_i(s-1)$.

This equations (really there are three ones) are selfconsistent and have the following unique solution
$$
b_i=\sqrt {{\{s-q+i\}\{p+2q-s+2-i\}\{p+q-s+1-i\} \over \{1\}^2\{p+2q-s+3-2i\}\{p+2q-s+2-2i\}}}
$$
$$
a_i=\sqrt {{\{i-1\}\{p+q+3-i\}\{q+2-i\} \over \{1\}^2\{p+2q-s+3-2i\}\{p+2q-s+4-2i\}}}
$$
We would like to notice that additional condition $a_i=0$ satisfies automatically.

\section{Different aproach and numerous equations of equivalence}

In this section we eould like to demondtrate some other one way for obtaining the results above.
If take into account these results to all relations below may be considered as numeous nontrivial equation of equivalence. The proving them directly is not a simple problem.
After multiplication (\ref{GRA}) on its transposes from the left we come to equation
\be
O^{s,k})(\lambda_q^{s+1,k})^2(O^{s,k})^{-1}=L^TL \label{PV}
\ee
From the explicit expression of $L$ matrix (\ref{YX}) we conclude immediately that different from zero elements of symmetrical $L^TL$ matrix are only on its main diagonal and 
on one step upper, down, right and left. The line of this matrix has the form
$$
(L^TL)_{i,i-1}= b_{i-1}a_i\sqrt{ \{k+2-i\}\{p+s-k+2-i\}}  
$$
$$ 
(L^TL)_{i,i}=a_i^2\{p+s-k+2-i\}+b_i^2\{k+1-i\}
$$
$$
(L^TL)_{i,i+1}= b_ia_{i+1}\sqrt{ \{k+1-i\}\{p+s-k+1-i\}}     
$$
In the relation (\ref{PV}) we know explicit expression for $(s+1)$ column of ortogonal matrix 
$O^{s,k}$ and all its proper values $(\lambda_q^{s+1,k})^2$. This give possibility to calculate all functios $a_i,b_i$ and find explicit expresion for matrix elements of $X^-_2=L$
generator of the second simple root.
The first line of equation (\ref{PV}) looks as
$$
\tilde l_1(a_1^2\{p+s-k+1\}+b_1^2\{k\})+ \tilde l_2 a_2b_1\{p+s-k\}=\tilde l_1{\{q+k-2s\}\over \{1\}}  
$$
where $\tilde l_i(s)\equiv l_i(s-1)$. Puting in the last equation $k=p+s$ and using equation
$b_1=-{l_1\over l_2} a_1$ we immidiately obtain:
$$
a_1=\sqrt{ {\{s+1\}\{q-s\}\over \{1\}^2 \{p+s+1\}}}, b_1=-\sqrt{ {\{p+q+1\}\{p\}\over \{1\}\{p+s\}\{p+s+1\}}}
$$
Substituting in the last equation $k=0$ we obtain $a_2$:
$$
a_2=\sqrt{ {\{s\}\{q-s+1\}\{p+s+1\}\over \{1\}^2\{p+s\} \{p+s-1\}}}, 
$$
Using the linear equation $b_2=-{l_2\over l_3}a_2$ we obtain
$$
b_2=-\sqrt{ {\{2\}\{p+q\}\{p-1\}\over \{1\}\{p+s-1\} \{p+s-2\}}}
$$
>From the second line of the equation (\ref{PV}) we obtain without of any diffecalties
$a_3,b_3$ and so on. The final result is the following (of course coinsedes with obtained in the previous section):
$$
a_i=\sqrt{ {\{s+2-i\}\{q-s+i-1\}\{p+s+3-i\}\over \{1\}^2\{p+s+3-2i\} \{p+s+2-2i\}}},\quad b_i=-\sqrt{ {\{i\}\{p+q+2-i\}\{p+1-i\}\over \{1\}^2\{p+s+3-2i\} \{p+s+4-2i\}}}
$$

\section{Outlook}

The results of the present paper give possibility to find (in principle, not to give explicit 
expressions) the generators of the simple roots for all quantum algebras of arbitrary rank with symmetrical Caratan matrices,i.e. for $A_n,D_n,E_6,E_7,E_8$ series.

Let us consider two first simple roots of the Dynkin diagramm. From Weyl formula we know the "spectral structure" of the representation diagram of the "big algebra" with respect to theirreducible representations of the algebra of the second rank connected with the simple roots $X^{\pm}_{1,2}$. In other words we know what irreducible representations $(p,q)$ of $X^{\pm}_{1,2}$ algebra are connected with each point of representation diagram. Let the simple root $X^{\pm}_1$ be connected with the $p$ lines $X^{\pm}_2$ with $q$ ones. With the help of inverse ortogonal rotation it is possible to transform the second simple root to diagonal form passing it generators to form of $p$ lines (i.e. to "diagonal form"). After this it is necessary to repeat all calculations with respect to second rank algebra connected with $X^{\pm}_{2,3}$ simple roots. Spectral structure also is known (from Weyl formula)
and so we have explicit form of the primitve matrix elements and ortogonal matrices in each point of the representation diagram. This procedure may be continued up to the last simple root
of corresponding Dynkin lattice of the "big algebra". This is the way how to construct the explicit expressions for generators of irreducible representations of the "big algebra" 
if all rotations angles of the used algebras of second rank are known. This exactly what is necessary for calculations of matrix elements of $A_n,D_n,E_6,E_7,E_8$ series.

\end{document}